\documentclass[10pt,twocolumn,letterpaper]{article}

\usepackage{cvpr}              % To produce the CAMERA-READY version

\usepackage{graphicx}
\usepackage{amsmath}
\usepackage{amssymb}
\usepackage{booktabs}
\usepackage[accsupp]{axessibility}
\usepackage{amsmath}
\newtheorem{theorem}{Theorem}
\newtheorem{lemma}{Lemma}

\usepackage[colorlinks,linkcolor=blue]{hyperref}

\usepackage[capitalize]{cleveref}
\crefname{section}{Sec.}{Secs.}
\Crefname{section}{Section}{Sections}
\Crefname{table}{Table}{Tables}
\crefname{table}{Tab.}{Tabs.}

\def\cvprPaperID{15} % *** Enter the CVPR Paper ID here
\def\confName{CVPR}
\def\confYear{2022}

\begin{document}

\title{Unsupervised Domain Adaptation for Cardiac Segmentation: Towards Structure Mutual Information Maximization}

\author{Changjie Lu, Shen Zheng, Gaurav Gupta\\
College of Science and Technology, Wenzhou-Kean University\\
Wenzhou, China\\
{\tt\small lucha,zhengsh,ggupta@kean.edu}
}
\maketitle

\begin{abstract}

Unsupervised domain adaptation approaches have recently succeeded in various medical image segmentation tasks. The reported works often tackle the domain shift problem by aligning the domain-invariant features and minimizing the domain-specific discrepancies. That strategy works well when the difference between a specific domain and between different domains is slight. However, the generalization ability of these models on diverse imaging modalities remains a significant challenge. This paper introduces UDA-VAE++, an unsupervised domain adaptation framework for cardiac segmentation. We show that the evidence lower bound of VAE-based domain adaptation penalizes the mutual information between the image and the latent space, including the structure information that segmentation relies on. Maximizing the structure mutual information removes this penalty on the structure information, while the remaining information is still compressed. To estimate this term, we develop a novel Structure Mutual Information Estimation (SMIE) block with a global estimator, a local estimator, and a prior information matching estimator to maximize the mutual information between the reconstruction and segmentation tasks. Specifically, we design a novel sequential reparameterization scheme that enables information flow and variance correction from the low-resolution latent space to the high-resolution latent space. Comprehensive experiments on benchmark cardiac segmentation datasets demonstrate that our model outperforms previous state-of-the-art qualitatively and quantitatively. The code is available at \href{https://github.com/LOUEY233/Toward-Mutual-Information}{https://github.com/LOUEY233/Toward-Mutual-Information}

\end{abstract}

\section{Introduction}

Deep learning-based methods have recently achieved promising results on various medical image processing tasks, such as detection \cite{liu2017detecting, yan2019mulan} and segmentation \cite{ronneberger2015u, dou20173d}. Indeed, deep learning approaches can generalize effectively when the training and testing images are from the same modality (i.e., same distribution), approaching or surpassing human-level performance.
 
\begin{figure}[t]
    \centering
    \includegraphics[width=1.55cm]{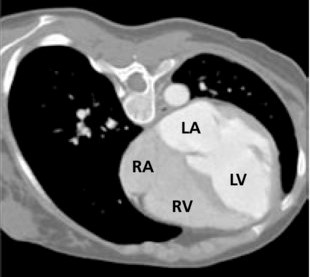}
    \includegraphics[width=1.7cm]{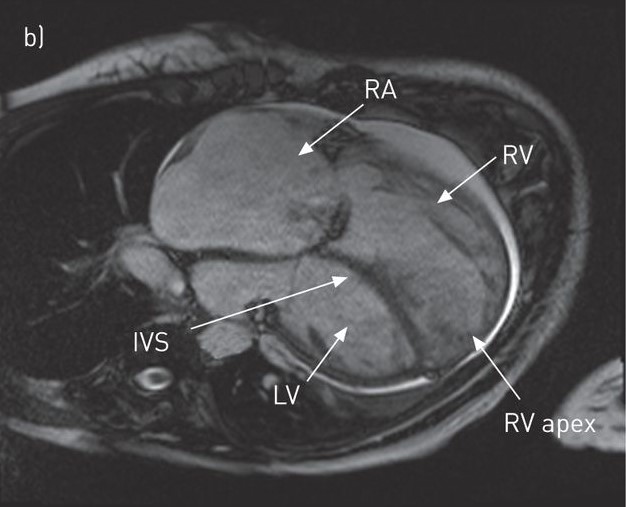}
    \includegraphics[width=1.575cm]{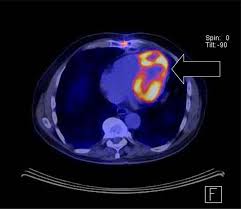}
    \includegraphics[width=1.7cm]{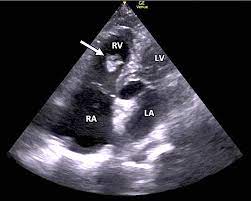}
    \caption{Four types of Cardiac Imaging. From left to right: Computerized Tomography (CT), Magnetic Resonance Imaging (MRI), Positron Emission Tomography (PET), and Ultrasound (US).\cite{MRICT} Each type has significant difference with others in terms of color, contrast, structure, artifacts, and the edge information.}
    \label{Showcase}
\end{figure}

However, some researchers \cite{kalogeiton2016analysing, tommasi2016learning} have shown that well-trained models do not perform well when the testing images come from a different statistical distribution from the training images. This domain shift problem is common in real-world medical diagnosis since medical images at various steps of the clinical procedure are often obtained with different physical properties \cite{dou2018unsupervised}. For instance, Magnetic Resonance Imaging (MRI) and Computed Tomography (CT) play complementary roles in cardiac disease diagnosis while also exhibiting different appearances (See Fig. \ref{Showcase}). That difference post challenges for analyzing the MRI and CT images in clinical diagnosis. 

One plausible solution is to obtain manual annotations for both the MRI and the CT images from medical experts. However, such a procedure is prohibitively time-consuming. (e.g., manual cardiac operations from MRI/CT consumes 2-4 hours \cite{zhuang2013challenges}). Unsupervised Domain adaptation (UDA), which automatically transfers knowledge from the source domain to the target domain (e.g., MRI to CT) without paired images, is an interesting idea.

For UDA with medical image segmentation, the source medical image with the ground truth segmentation is denoted as the source domain, whereas the target medical image without the ground truth segmentation is referred to as the target domain. Generally, the reported works such as \cite{dou2018unsupervised, dou2019pnp} align the source domain and the target domain by learning the domain-invariant features and minimizing the domain-specific discrepancies.

One popular research direction is to combine UDA with a GAN-based strategy, as GAN \cite{goodfellow2014generative} and its derivatives \cite{zhu2017unpaired, isola2017image} have exhibited remarkable unsupervised domain adaption ability. In the GAN-based UDA approach, the domain-invariant latent space features can be implicitly learned via adversarial learning during the min-max game between the generator and the discriminator. GAN-based approaches \cite{zhang2018translating, chen2020unsupervised, liu2020duda} have recently gained widespread acceptance in medical image analysis, outperforming prior Convolutional Neural Network (CNN) methods such as \cite{dou2018unsupervised, dou2019pnp} on cardiac segmentation tasks \cite{zhuang2016multi, zhang2018translating}. However, when we have a dataset (e.g., \cite{zhuang2018multivariate}) with extremely diverse imaging modalities and scanning methods, GAN-based approaches often fail to converge to the Nash Equilibrium \cite{heusel2017gans, karras2019style, karras2020analyzing, wu2021unsupervised}.

Recently, researchers \cite{ouyang2019data, gu2022few, wu2020cf, wu2021unsupervised} in UDA with medical imaging have turned to Variational Autoencoder (VAE) \cite{kingma2013auto} as the backbone due to their training stability at domain adaptation tasks at diverse imaging modalities \cite{purushotham2016variational, wu2021unsupervised} and their ability to handle scarce data in the target domain \cite{ouyang2019data, gu2022few}. These VAE-based methods usually perform posterior inference for the latent space variables using the normal distribution. That property allows it to consistently bridge two domains (i.e., source and target domain) towards standard and parameterized latent space variables \cite{wu2021unsupervised}.

Despite VAE-based methods' excellent domain adaptation ability at challenging benchmark cardiac segmentation datasets (e.g., \cite{zhuang2018multivariate}), two crucial factors restrain their learning capability. Firstly, VAE-based methods like UDA-VAE \cite{wu2021unsupervised} introduce a separate image reconstruction stage, aiming to regularize the latent space towards normal distribution. Although this strategy could explicitly minimize the domain discrepancy, the information from the reconstructed output cannot be directly delivered to the segmentation. Moreover, the KL term of the evidence lower bound compresses the latent space without distinguishing the structure information that segmentation needs. Secondly, VAE-based approaches like CFDNet \cite{wu2020cf} utilize parallel reparameterization for latent space with different resolutions. The separation of low-resolution latent space and high-resolution latent space in U-Net-like architecture will potentially exaggerate the domain shift problem \cite{yan2019domain} and, therefore, degrade the model performance.

In this work, we propose a new framework, dubbed UDA-VAE++, that can well address unsupervised domain adaptation in cardiac image segmentation with diverse imaging modalities. Firstly, we leverage a U-Net backbone to extract the multi-scale features from unpaired images from the source and the target domain. The output at each encoder stair enters variational reasoning, followed by our sequential reparametrization design. That sequential design enables the network to transfer knowledge from low-resolution latent space to high-resolution latent space and constrains the encoded output according to standard normal distribution. A segmentation block follows the reparametrization operation at each level, and the segmentation output will be passed into a reconstruction block. Finally, we conduct mutual information (global, local, prior) estimation and maximization for the segmentation output and the reconstruction, which keeps the structure information in the latent space from being compressed by the evidence lower bound.

The main contributions of this paper are highlighted as follows:
\begin{itemize}
    \item We show that the evidence lower bound of VAE-based domain adaptation penalizes the mutual information $I(x;z)$ between the image and the latent space, and that the structure mutual information term reduces this penalty to $I(x;z\mid\hat{y})$, the information that the segmentation does not explain.
    
    \item We design a novel, plug-and-play style, Structure Mutual Information Estimation (SMIE) block. This design enables an efficient mutual information estimate for the reconstruction output and the segmentation output, making the reconstruction and segmentation tasks mutually beneficial.
    
    \item We convert parallel reparameterization to sequential reparameterization, allowing information flow and variance correction from the low-resolution latent space to the high-resolution latent space after variational reasoning.

    \item We conduct extensive experiments to demonstrate that the proposed method surpasses previous state-of-the-arts on benchmark cardiac segmentation datasets qualitatively and quantitatively.
\end{itemize}

\section{Related Work}

\subsection{Unsupervised Domain Adaptation}

Unsupervised Domain Adaptation (UDA) has been widely used for biomedical image segmentation tasks. The early works \cite{dou2018unsupervised} and \cite{dou2019pnp} leverage unsupervised domain adaptation with adversarial training for multi-modal biomedical image segmentation. Specifically, both papers utilize a plug-and-play domain adaptation module to align the features in the source and the target domain.

Due to the promising generalization ability of Generative Adversarial Network (GAN) \cite{goodfellow2014generative}, recent research has begun to incorporate GAN in UDA for biomedical image segmentation. For example, \cite{zhang2018translating} utilizes CycleGAN \cite{zhu2017unpaired} with a shape-consistency loss to realize cross-domain translation between CT and MRI images. SIFA \cite{chen2020unsupervised} presents a synergistic domain alignment at both image-level and feature-level using the adversarial learning of CycleGAN to exploit domain-invariant characteristics. DUDA \cite{liu2020duda} further incorporates a cross-domain consistency loss to improve the segmentation performances.

Another faithful research direction is to use Variational Autoencoder (VAE) \cite{kingma2013auto}. That strategy is advantageous when there are few images in the target domain. For instance, \cite{ouyang2019data} follows the few-shot learning strategy, integrating a VAE-based feature prior to matching with adversarial learning to exploit the domain-invariant features. FUDA \cite{gu2022few} further incorporates Random Adaptive Instance Normalization to explore diverse target styles where there is only one unlabeled image in the target domain. The recent work CFDNet \cite{wu2020cf} proposes an effective metric, dubbed CF Distance, which enables explicit domain adaptation with image reconstruction and prior distribution matching. Another work UDA-VAE \cite{wu2021unsupervised} goes even further: it drives the latent space of the source and target domains towards a common, parameterized variational form following Gaussian Distribution.

Compared with previous UDA approaches, our method is the first that sequentially integrates multi-scale latent space features. That design enables our network to effectively minimize the domain-specific discrepancy according to the information flow from the low-resolution latent space to the high-resolution latent space.

\subsection{Mutual Information Neural Estimation}
Mutual Information Neural Estimation (MINE) is first introduced in \cite{belghazi2018mutual}, where the author utilizes gradient descent algorithms over neural networks to approximate the mutual information between continuous random variables. Based upon MINE, Deep InfoMax (DIM) \cite{hjelm2018learning} explores unsupervised visual representation learning by maximizing the mutual information for the network input and the encoded output under statistical constrain. A recent work \cite{chen2020structure} utilizes MINE to address the domain shift problem in unsupervised domain adaptation. Specifically, that paper integrates network predictions and local features into global features by simultaneously maximizing the mutual information.

Recently, MINE has been applied in biomedical image processing tasks. For example, based on MINE, \cite{ting2020multiview} maximizes the mutual information between source and fused images from Multiview 3-D Echocardiography. \cite{snaauw2022mutual} tackle the challenging unsupervised multimodal brain image segmentation task by estimating the mutual information using a lightweight convolutional neural network.

Different from previous MINE approaches, our framework is the first that conducts mutual information estimation and maximization with both image reconstruction and image segmentation. Our unique design enables image reconstruction and image segmentation to be mutually beneficial during model learning.

\begin{table}[t]
\setlength{\tabcolsep}{1mm}{
\centering{
\begin{tabular}{llll}
\toprule[1.0pt]
 & Symbols & Description                                         &  \\ \hline
 & $S$ & Source domain & \\
 & $T$ & Target domain & \\
 & $z$ & Latent variable & \\
 & $x$ & Input image data point & \\
 & $y$ & Ground truth segmentation & \\
 & $\hat{y}$ & Predicted segmentation & \\
 & $R$ & Reconstructed image & \\
 & $p_{D}(x,y)$ & Data distribution & \\
 & $p_{\theta}()$ & PDF of variables with parameter $\theta$ & \\
 & $q_{\phi}(z\mid x)$ & Encoder with parameter $\phi$ & \\
 & $q(z)$ & Aggregate posterior & \\
 & $\mathcal{L}_{D}$ & Domain distance between source and target & \\
 & $D_{KL}$ & KL Divergence & \\
 & $I$ & Mutual information & \\
 & $\mathcal{E}$ & Reconstruction and segmentation error & \\
 & $\lambda$ & Weight of the structure mutual information & \\
\bottomrule[1.0pt]

\end{tabular}}
}
\caption{Preliminary for Important Symbols}
\end{table}
\begin{figure*}[t]
    \centering
    \includegraphics[width=15cm]{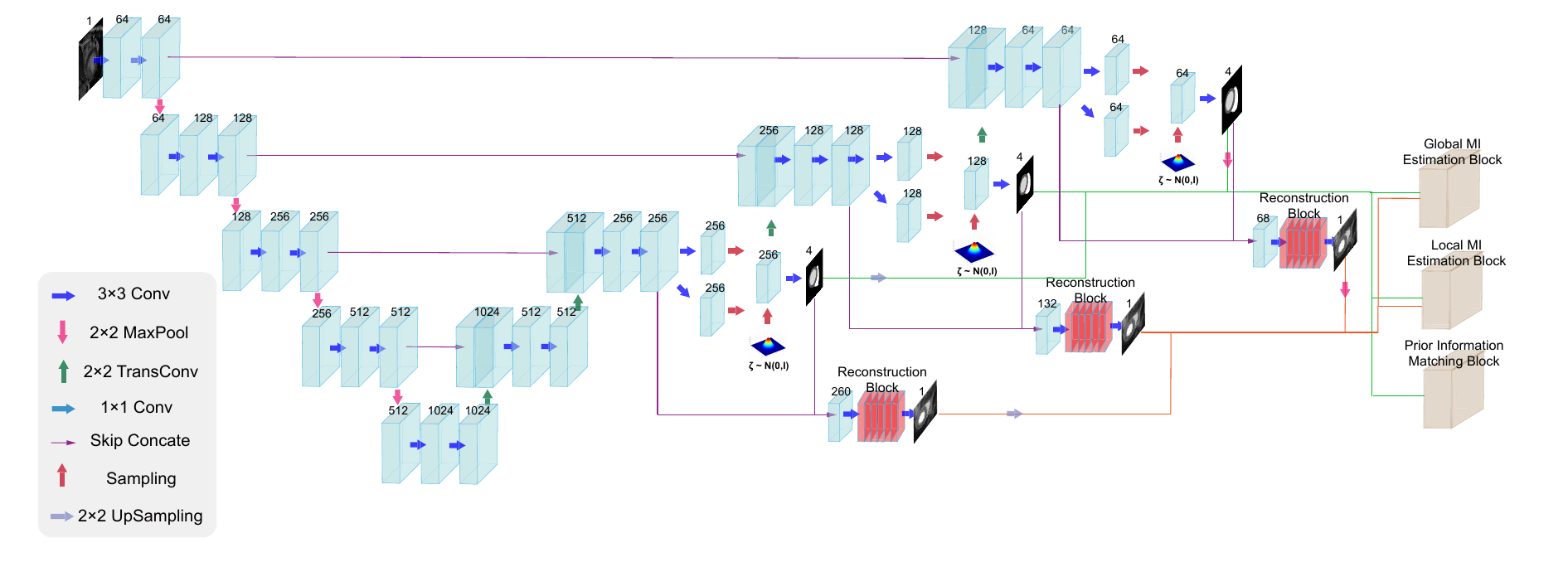}
    \caption{The Model Architecture of UDA-VAE++. The backbone of UDA-VAE++ is U-Net (blue boxes) with three scales of variational blocks. The green line refers to the concatenation of the segmentation output, whereas the orange line indicates the concatenation of the reconstruction output. The reconstruction blocks (red boxes) contain seven convolution layers. The grey box refers to the MI estimation block detailed in Fig. \ref{MINE}}
    \label{Preliminary}
\end{figure*}

\section{Methodology}
In this section, we will discuss our UDA-VAE++ workflow, explain the proposed structure mutual information estimation block, and display the loss functions.
\subsection{UDA-VAE++ Model Workflow}
\begin{figure}[t]
    \centering
    \includegraphics[width=10cm]{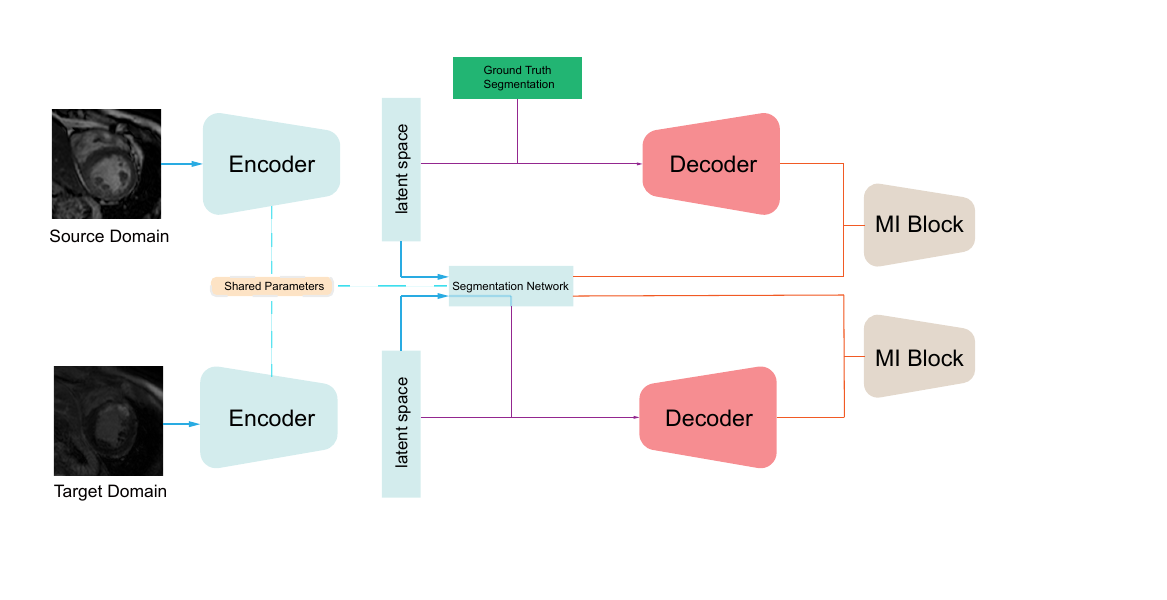}
    \caption{The workflow for Unsupervised Domain Adaptation. The image from the source and target domain will first be encoded in the shared parameters downsampling part of the U-Net backbone. Next, each scale output will go through the same segmentation network. In the source domain, the ground truth segmentation masks combining the variables in latent space will be reconstructed by the upsampling part of U-Net. The MI block will maximize the mutual information of the segmentation output and the reconstruction output.}
    \label{UDAworkflow}
\end{figure}

As shown in Fig. \ref{Preliminary}, we use U-Net \cite{ronneberger2015u} as our backbone due to its remarkable success in medical image segmentation. Firstly, The network performs four downsamplings. Each of the downsampling operations uses two convolutional layers. Secondly, the network uses upsampling symmetrically with skip connection. We then obtain a multi-scale encoding output with channels of 256, 128, 64, and image sizes of 40$\times$40, 80$\times$80, 160$\times$160, respectively. Each encoding output will be followed by variational reasoning \cite{kingma2013auto, kingma2019introduction}. Using the reparameterization trick \cite{kingma2013auto} with the latent mean variable, the latent log variance variable, and the standard normal distribution, we obtain three latent variables $z_{1},z_{2},z_{3}$, which we denote jointly by $z$. After that, We use a single convolutional layer to obtain the predicted segmentation $\hat{y}$.

Finally, we leverage a fully convolutional network with 7 layers for image reconstruction. The input for the source domain includes the ground truth segmentation $y$ and the latent variable $z$, whereas the input for the target domain is the predicted segmentation $\hat{y}$.

\subsection{Structure Mutual Information Estimation}
In this subsection, we aim to estimate the mutual information between the segmentation outcome $\hat{y}$ and the reconstruction output $R$ in the source and target domains. The mutual information can be formulated as:

\begin{equation}
     I\left(\hat{y} ; R\right)=D_{K L}\left(\mathbb{P}_{\hat{y} R} \| \mathbb{P}_{\hat{y}} \otimes \mathbb{P}_{R}\right)
\end{equation}
The KL divergence between joint distribution $\mathbb{P}_{\hat{y} R}$ and marginal distribution $\mathbb{P}_{\hat{y}}\otimes \mathbb{P}_{R}$ can be written as its dual representation\cite{donsker1975asymptotic} as below:
\begin{equation}
    D_{K L}(\mathbb{P}_{\hat{y} R} \| \mathbb{P}_{\hat{y}} \otimes \mathbb{P}_{R})=\sup _{T: \Omega \rightarrow \mathbb{R}}( \mathbb{E}_{\mathbb{P}_{\hat{y} R}}[T]-\log \left(\mathbb{E}_{\mathbb{P}_{\hat{y}} \otimes \mathbb{P}_{R}}\left[e^{T}\right]\right))
\end{equation}
where $\Omega$ is the domain of $(\hat{y}, R)$ and the supremum is taken over all functions $T$; in practice, $T$ is a neural network.

Inspired by \cite{hjelm2018learning}, we are interested in automatically maximizing the mutual information rather than manually obtaining the exact value for mutual information. The mutual information maximization process can be formulated as:

\begin{equation}
    \widehat{\mathcal{I}}\left(\hat{y} ; R\right) = \mathbb{E}_{\mathbb{P}_{\hat{y} R}}\left[-\operatorname{sp}\left(-T\left(\hat{y}, R\right)\right)\right]-\mathbb{E}_{\mathbb{P}_{\hat{y}} \otimes \mathbb{P}_{R}}\left[\operatorname{sp}\left(T\left(\hat{y}, R^{\prime}\right)\right)\right]
\label{mi}
\end{equation}
where $R^{\prime}$ is drawn from the marginal $\mathbb{P}_{R}$, i.e., the reconstruction of another image in the batch, and $\operatorname{sp}(z)=\log \left(1+e^{z}\right)$ is the softplus function.

The next step is to estimate the joint and marginal distribution of $\hat{y}$ and $R$ using contrastive learning. First, we design three estimators in the MI block \cite{hjelm2018learning}. The original paired $R$ and $\hat{y}$ serve as the anchor and the positive point, respectively. We then shuffle $R$ randomly to obtain the negative point. To fuse the data together, we upsample the 40$\times$40 feature map and downsample the 160$\times$160 feature map. Before entering the estimator block, the anchor and negative point will go through two convolutional layers, whereas the positive point will go through three convolutional layers. 

For the Global MI Estimation block, we concatenate the positive points with anchor and negative points, pushing the anchor away from the negative points and pulling the anchor towards the positive point. For the Local MI Estimation block, we extract the high-level semantics using fully connected layers. Next, we concatenate the semantic information with the positive point to acquire the locality information, followed by two convolutional layers for contrastive learning. 

Finally, motivated by \cite{hjelm2018learning, chen2020structure}, we adopt the prior matching \cite{makhzani2015adversarial} strategy to constrain the visual representations according to standard normal distribution. Specifically, in the prior information estimation block, the positive point will go through fully connected layers and output the prior information. 

\begin{figure*}[t]
    \centering
    \includegraphics[width=12cm]{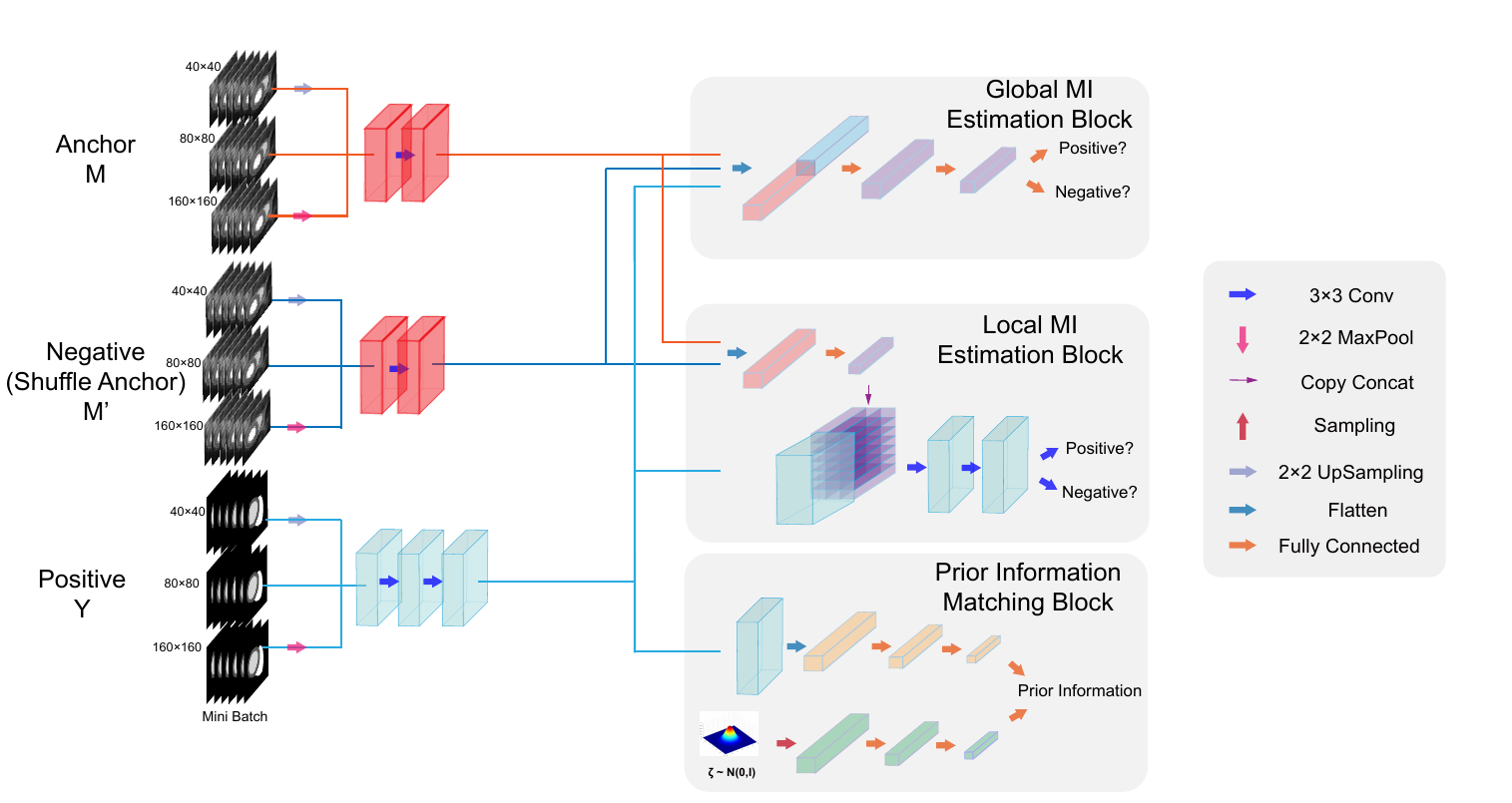}
    \caption{The architecture of Structure Mutual Information Estimation (SMIE) block. We use the reconstruction image as anchor, the shuffle reconstructed images as negative points, and the segmented image as positive points. The Global \& Local Mutual Information (MI) Estimation Block follows contrastive learning schemes to maximize mutual information, whereas the prior information matching block align the positive point with the standard normal distribution. Finally, the sum of the outputs score from these three blocks serves as the loss function for $\mathcal{L}_{MI}$.}
    \label{MINE}
\end{figure*}

\begin{figure}[t]
    \centering
    \includegraphics[width=8cm]{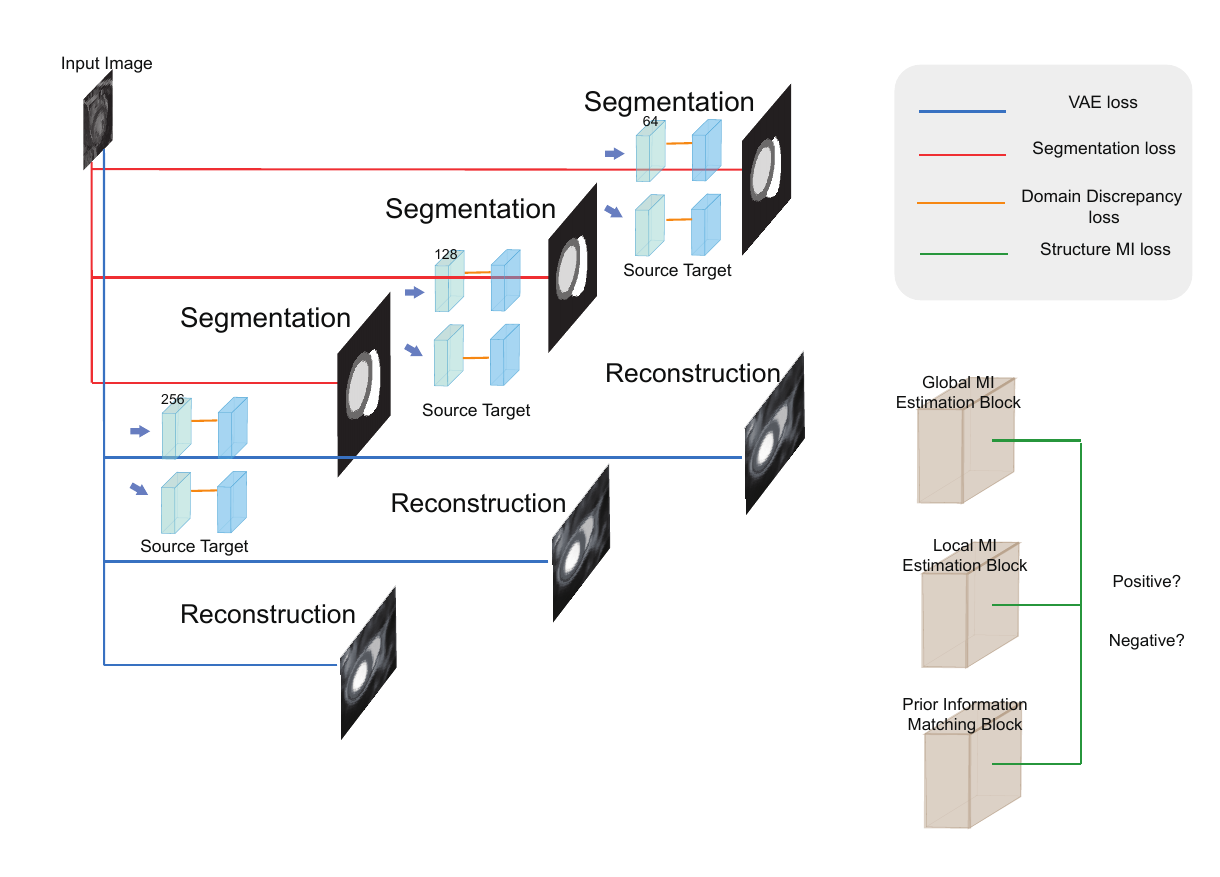}
    \caption{The loss function of the proposed method. The blue line refers to the reconstruction loss. The red line indicates the segmentation loss between the predicted segmentation and ground truth segmentation. The orange line illustrates the domain discrepancy loss in the latent space. The green line refers to the structure mutual information loss.}
    \label{Loss}
\end{figure}

\subsection{Loss function}
We first analyze the variational objective of UDA-VAE~\cite{wu2021unsupervised} and then derive our loss. We state the results for the source domain and drop the subscript $S$; the target domain is identical with $y$ replaced by the prediction $\hat{y}$. Let $p_{D}(x,y)$ be the data distribution, $q_{\phi}(z\mid x)$ the encoder, $q(x,y,z)=p_{D}(x,y)\,q_{\phi}(z\mid x)$, and $q(z)=\mathbb{E}_{p_{D}}[q_{\phi}(z\mid x)]$ the aggregate posterior. All mutual informations are computed under $q$. The generative model is $p_{\theta}(x,y,z)=p(z)\,p_{\theta}(y\mid z)\,p_{\theta}(x\mid y,z)$ with $p(z)=\mathcal{N}(0,\mathbf{I})$, and the segmentation $\hat{y}$ is computed from $z$. UDA-VAE maximizes the lower bound
\begin{equation}
\begin{aligned}
\mathrm{LB}(x,y)= & -D_{KL}\left(q_{\phi}(z\mid x)\,\|\,p(z)\right)\\
& +\mathbb{E}_{q_{\phi}(z\mid x)}[\log p_{\theta}(x\mid y,z)]\\
& +\mathbb{E}_{q_{\phi}(z\mid x)}[\log p_{\theta}(y\mid z)].
\end{aligned}
\label{lb}
\end{equation}
We write $\mathcal{E}=-\mathbb{E}_{q}[\log p_{\theta}(x\mid y,z)+\log p_{\theta}(y\mid z)]$ for the reconstruction and segmentation error.

Taking the expectation over the data, the lower bound decomposes as
\begin{equation}
\mathbb{E}_{p_{D}}[\mathrm{LB}]=-\mathcal{E}-I(x;z)-D_{KL}\left(q(z)\,\|\,p(z)\right),
\label{elbo_mi}
\end{equation}
and adding the structure mutual information with weight $\lambda\geq 0$ gives
\begin{equation}
\begin{aligned}
&\mathbb{E}_{p_{D}}[\mathrm{LB}]+\lambda I(x;\hat{y})\\
=&-\mathcal{E}-D_{KL}\left(q(z)\,\|\,p(z)\right)-I(x;z\mid\hat{y})-(1-\lambda)\,I(x;\hat{y})
\end{aligned}
\label{loss_fun}
\end{equation}
(Theorem~\ref{thm:main} in the supplementary material). Eq.~\ref{elbo_mi} shows that the compression term of the evidence lower bound penalizes $I(x;z)$, the information that the latent space carries about the image~\cite{hoffman2016elbo}. This penalty does not distinguish the structure information from the rest, since $I(x;z)=I(x;\hat{y})+I(x;z\mid\hat{y})$. Compressing $I(x;z\mid\hat{y})$, which includes the modality-specific appearance, helps domain adaptation, but compressing $I(x;\hat{y})$ removes the structure information that segmentation needs. Adding the structure mutual information with weight $\lambda$ changes the penalty on the structure information to $(1-\lambda)I(x;\hat{y})$ and keeps the penalty on the remaining information (Eq.~\ref{loss_fun}).

The SMIE block maximizes $I(\hat{y};R)$ as a surrogate for $I(x;\hat{y})$ (Lemma~\ref{lem:bridge} in the supplementary material).

Our loss function (Fig. \ref{Loss}) contains structure mutual information estimation loss $\mathcal{L}_{MI}$ (the term $\lambda I(x;\hat{y})$ in Eq.~\ref{loss_fun}), reconstruction loss $\mathcal{L}_{recon}$ (the first two terms of Eq.~\ref{lb}), segmentation loss $\mathcal{L}_{seg}$ (the last term of Eq.~\ref{lb}), and domain discrepancy loss $\mathcal{L}_{D}$, which we minimize explicitly as the latent space obeys normal distribution.

\subsubsection{Reconstruction Loss}
The reconstruction loss corresponds to the first two terms of Eq.~\ref{lb}. The first term is the KL divergence between the encoder $q_{\phi}(z|x)$ and the prior distribution $p(z)=\mathcal{N}(0,\mathbf{I})$, where $\mathbf{I}$ is the identity matrix. The encoder $q_{\phi}(z|x)$ performs variational reasoning upon $u$ and $\sigma^{2}$ to approximate $0$ and $\mathbf{I}$, respectively. With the reparameterization trick\cite{vae}(red arrows in Fig. \ref{Preliminary}), we can get:
\begin{equation}
    D_{K L}\left(q_{\phi}(z|x) \| p(z)\right)=\frac{1}{2} \sum_{l=1}^{n}\left(\sigma_{l}^{2}+u_{l}^{2}-\log \sigma_{l}^{2}-1\right)
\end{equation}
where $n$ is the dimension of $z$. The second term is the log-likelihood of $x$ given $y$ and $z$. It is calculated by the cross entropy loss between the input $x$ and the reconstruction output $R$:
\begin{equation}
    \mathcal{L}_{ce} = -(x \log (R)+(1-x) \log (1-R))
\end{equation}
Finally, we get the reconstruction loss:
\begin{equation}
    \mathcal{L}_{recon} = D_{KL} + \mathcal{L}_{ce}
\end{equation}

\subsubsection{Segmentation Loss}
The segmentation loss helps us minimize the loss between the predicted segmentation $\hat{y}$ and the ground truth segmentation $y$. We apply cross-entropy loss over the classes $c$, which is formulated as below:
\begin{equation}
    \mathcal{L}_{seg} = -\sum_{c} y_{c} \log \hat{y}_{c}
\end{equation}

\subsubsection{Domain Discrepancy Loss}

The Domain Discrepancy Loss helps reduce the domain discrepancy between the source and the target domain in the latent space. In the UDA-VAE framework, \cite{wu2021unsupervised} has proved that optimizing the distance explicitly would have better accuracy than adversarial training. As the latent space is regularized into a standard normal distribution, we can calculate the distance analytically. The Domain Discrepancy Loss is formulated as below:

\begin{equation}
\begin{aligned}
    & \mathcal{L}_{D} = \sum_{l=1}^{n}\int [q_{S}(z_{l})-q_{T}(z_{l})]^{2}dz_{l} \\
    & \approx \frac{1}{M^{2}} \sum_{i=1}^{M} \sum_{j=1}^{M}\left[k\left(x_{S_{i}}, x_{S_{j}}\right)+k\left(x_{T_{i}}, x_{T_{j}}\right)-2 k\left(x_{S_{i}}, x_{T_{j}}\right)\right]
\end{aligned}
\end{equation}
where $z_{l}$ is the $l$-th element of $z$, $q_{S}(z_{l})$ and $q_{T}(z_{l})$ are the marginals of the aggregate posterior $q(z)$ in the source and target domains, $M$ is the batch size, and $i,j$ index the samples in one batch. As the posterior $q_{\phi}(z\mid x)$ is Gaussian with mean $u$ and variance $\sigma^{2}$, the kernel function $k$ is:
\begin{equation}
    k\left(x_{S_{i}}, x_{T_{j}}\right) = \sum_{l=1}^{n}(2\pi)^{-\frac{1}{2}} e^{-\frac{1}{2}[\frac{(u_{S_{i},l}-u_{T_{j},l})^{2}}{\sigma_{S_{i},l}^{2}+\sigma_{T_{j},l}^{2}}+\log(\sigma_{S_{i},l}^{2}+\sigma_{T_{j},l}^{2})]}
\end{equation}

\subsubsection{Structure Mutual Information Loss}
As discussed in Eq.~\ref{mi}, we design a contrastive learning framework to estimate the joint and marginal distribution of $\hat{y}$ and $R$.
To maximize $I(\hat{y} ; R)$, we design a global MI estimation block, a local MI estimation block, and a prior information matching block.
\begin{equation}
    \mathcal{L}_{MI} = -(\alpha \widehat{\mathcal{I}}_{Global}(\hat{y} ; R)+ \beta \widehat{\mathcal{I}}_{Local}(\hat{y} ; R)+\gamma \widehat{\mathcal{I}}_{Prior})
\end{equation}
where $\alpha, \beta, \gamma$ are set as 0.5, 1.0, 0.1. $\widehat{\mathcal{I}}_{Prior} = \mathbb{E}[\log D_{\psi}(v)]+\mathbb{E}[\log(1-D_{\psi}(\hat{y}))]$, where $v\sim\mathcal{N}(0,\mathbf{I})$ and $D_{\psi}$ is the discriminator of the prior information matching block.
\subsubsection{Total Loss}
The total loss is defined as:
\begin{equation}
\begin{aligned}
    \mathcal{L}_{total} &= (c1\mathcal{L}_{recon}+c2\mathcal{L}_{seg}+c3\mathcal{L}_{MI})_{source} \\
    &+ (c1\mathcal{L}_{recon}+
    c2\mathcal{L}_{seg}+c3\mathcal{L}_{MI})_{target} \\
    &+ c4\mathcal{L}_{D}
\end{aligned}
\end{equation}
where c1, c2, c3, c4 are empirically set as 1e-2, 1, 1e-1, 1e-5, respectively, and c3 corresponds to $\lambda$ in Eq.~\ref{loss_fun}.

\section{Experiments}
\begin{table}[t]
\setlength{\tabcolsep}{1mm}{
\begin{tabular}{lllllllll}
\hline
\multicolumn{6}{c}{Model Components}                  & \multicolumn{3}{c}{Dice ($\%$)}                         \\ \hline
Base & SR & Att & Global & Local & Prior & MYO            & LV             & RV             \\ \hline
\checkmark      &    &             &        &       &       & 68.42          & 84.41          & 72.59          \\
\checkmark      & \checkmark  &             &        &       &       & 68.56          & 84.07          & 74.06          \\
\checkmark       & \checkmark  & \checkmark           &        &       &       & 68.30          & 84.91          & 74.72          \\
\checkmark       & \checkmark  & \checkmark           & \checkmark      &       &       & 69.25          & 84.70          & 75.63          \\
\checkmark       & \checkmark  & \checkmark           & \checkmark      & \checkmark     &       & 68.49          & 87.50          & \textbf{77.37} \\ \hline
\checkmark       & \checkmark  &             & \checkmark      & \checkmark     & \checkmark     & \textbf{70.75} & \textbf{88.64} & 75.82          \\
\checkmark       & \checkmark  & \checkmark           & \checkmark      & \checkmark     & \checkmark     & \textcolor{blue}{69.81}        & \textcolor{blue}{87.54}        & \textcolor{blue}{77.13}         \\
 \hline
\end{tabular}}
\caption{The Ablations of model components for MS-CMRSeg Dataset from \textbf{bSSFP to LGE-MRI}. Base: UDA-VAE \cite{wu2021unsupervised}. SR: Sequential Reparameterization. Att: Attention. Global: Global MI Estimation Block. Local: Local MI Estimation Block. Prior: Prior Matching. The best score for UDA from bSSFP to LGE-MRI is in \textbf{bold} while the second-best score is in \textcolor{blue}{blue}.}
\label{ablation}
\end{table}

\begin{table}[t]
\setlength{\tabcolsep}{1mm}{
\begin{tabular}{lllllll}
\hline
 & \multicolumn{3}{c}{Dice ($\%$)}                                                                                                                                              & \multicolumn{3}{c}{ASSD (mm)}                                                                                                                                               \\ \hline
                & MYO                                                   & LV                                                    & RV                                                    & MYO                                                   & LV                                                    & RV                                                    \\ \hline
NoAdapt
& 14.50                                                 & 34.51                                                 & 31.10                                                 & 21.6                                                     & 11.3                                                     & 14.5                                            \\

CFDNet \cite{wu2020cf}          & \begin{tabular}[c]{@{}l@{}}64.21\\ \end{tabular} & \begin{tabular}[c]{@{}l@{}}81.39\\ \end{tabular} & \begin{tabular}[c]{@{}l@{}}72.30\\ \end{tabular} & \begin{tabular}[c]{@{}l@{}}2.81\\ \end{tabular} & \begin{tabular}[c]{@{}l@{}}3.41\\ \end{tabular} & \begin{tabular}[c]{@{}l@{}}4.91\\ \end{tabular} \\
SIFA \cite{chen2020unsupervised}          & \begin{tabular}[c]{@{}l@{}}67.69\\ \end{tabular} & \begin{tabular}[c]{@{}l@{}}83.31\\ \end{tabular} & \begin{tabular}[c]{@{}l@{}}\textbf{79.04}\\ \end{tabular} & \begin{tabular}[c]{@{}l@{}}2.56\\ \end{tabular} & \begin{tabular}[c]{@{}l@{}}3.44\\ \end{tabular} & \begin{tabular}[c]{@{}l@{}}\textbf{2.13}\\ \end{tabular} \\
UDA-VAE \cite{wu2021unsupervised}        & \begin{tabular}[c]{@{}l@{}}68.42\\ \end{tabular} & \begin{tabular}[c]{@{}l@{}}84.41\\ \end{tabular} & \begin{tabular}[c]{@{}l@{}}72.59\\ \end{tabular} & \begin{tabular}[c]{@{}l@{}}2.39\\ \end{tabular} & \begin{tabular}[c]{@{}l@{}}2.59\\ \end{tabular} & \begin{tabular}[c]{@{}l@{}}3.97\\ \end{tabular} \\ \hline
UDA-VAE++       & \begin{tabular}[c]{@{}l@{}}\textbf{70.75}\\ \end{tabular} & \begin{tabular}[c]{@{}l@{}}\textbf{88.64}\\ \end{tabular} & \begin{tabular}[c]{@{}l@{}}75.82\\ \end{tabular} & \begin{tabular}[c]{@{}l@{}}\textbf{2.02}\\ \end{tabular} & \begin{tabular}[c]{@{}l@{}}\textbf{2.27}\\ \end{tabular} & \begin{tabular}[c]{@{}l@{}}3.62\\ \end{tabular} \\
\hline
\end{tabular}}
\caption{Unsupervised Domain Adaptation for MS-CMRSeg Dataset from \textbf{bSSFP to LGE-MRI}. The best score for Dice$\uparrow$ and ASSD$\downarrow$ are in \textbf{bold}.}
\label{CT_MRI_Dice_IOU}
\end{table}

\begin{table}[t]
\setlength{\tabcolsep}{1mm}{
\begin{tabular}{lllllll}
\hline
         & \multicolumn{3}{c}{Dice ($\%$)}                                                                                                                                              & \multicolumn{3}{c}{ASSD (mm)}                                                                                                                                               \\ \hline
                         & MYO                                                   & LV                                                    & RV                                                    & MYO                                                   & LV                                                    & RV                                                    \\ \hline
NoAdapt & 12.32                                                 & 30.24                                                 & 37.25                                                 & 24.9                                                     &    10.4                                                  & 16.7                                           \\

CFDNet \cite{wu2020cf}                   & \begin{tabular}[c]{@{}l@{}}57.41\\ \end{tabular} & \begin{tabular}[c]{@{}l@{}}78.44\\ \end{tabular} & \begin{tabular}[c]{@{}l@{}}77.63\\ \end{tabular} & \begin{tabular}[c]{@{}l@{}}3.61\\ \end{tabular} & \begin{tabular}[c]{@{}l@{}}3.87\\ \end{tabular} & \begin{tabular}[c]{@{}l@{}}2.49\\ \end{tabular} \\
SIFA \cite{chen2020unsupervised}                  & \begin{tabular}[c]{@{}l@{}}60.89\\ \end{tabular} & \begin{tabular}[c]{@{}l@{}}79.32\\ \end{tabular} & \begin{tabular}[c]{@{}l@{}}\textbf{82.39}\\ \end{tabular} & \begin{tabular}[c]{@{}l@{}}3.44\\ \end{tabular} & \begin{tabular}[c]{@{}l@{}}3.65\\ \end{tabular} & \begin{tabular}[c]{@{}l@{}}1.80\\ \end{tabular} \\
UDA-VAE \cite{wu2021unsupervised}                 & \begin{tabular}[c]{@{}l@{}}58.58\\ \end{tabular} & \begin{tabular}[c]{@{}l@{}}79.43\\ \end{tabular} & \begin{tabular}[c]{@{}l@{}}80.43\\ \end{tabular} & \begin{tabular}[c]{@{}l@{}}3.53\\ \end{tabular} & \begin{tabular}[c]{@{}l@{}}3.27\\ \end{tabular} & \begin{tabular}[c]{@{}l@{}}2.04\\ \end{tabular} \\ \hline
UDA-VAE++                & \begin{tabular}[c]{@{}l@{}}\textbf{68.74}\\ \end{tabular} & \begin{tabular}[c]{@{}l@{}}\textbf{85.08}\\ \end{tabular} & \begin{tabular}[c]{@{}l@{}}81.42\\ \end{tabular} & \begin{tabular}[c]{@{}l@{}}\textbf{2.34}\\ \end{tabular} & \begin{tabular}[c]{@{}l@{}}\textbf{2.61}\\ \end{tabular} & \begin{tabular}[c]{@{}l@{}}\textbf{1.71}\\ \end{tabular} \\
\hline
\end{tabular}}
\caption{Unsupervised Domain Adaptation for MS-CMRSeg Dataset from \textbf{LGE-MRI to bSSFP}. The best score for Dice $\uparrow$ and ASSD$\downarrow$ are in \textbf{bold}.}
\label{MRI_CT_Dice_IOU}
\end{table}

\begin{table*}[t]
\centering
\begin{tabular}{lllllllllll}
\hline
Methods      & \multicolumn{5}{c}{Dice ($\%$)}                                                      & \multicolumn{5}{c}{ASSD (mm)}                                                      \\ \hline
             & MYO           & LA            & LV            & RA            & RV            & MYO           & LA            & LV            & RA            & RV            \\ \hline
NoAdapt      & 0.08        & 3.08          & 0.00          & 0.74         & 23.9          & --            & --            & --            & --            & --            \\
PnP-AdaNet \cite{dou2019pnp}      & 32.7          & 49.7          & 48.4          & 62.4          & 44.2          & 6.89          & 22.6          & 9.56          & 20.7          & 20.0          \\
SIFA \cite{chen2020unsupervised}        & 37.1          & 65.7          & 61.2          & 51.9          & 18.5          & 11.8          & 5.47          & 16.0          & 14.7          & 21.6          \\
UDA-VAE \cite{wu2021unsupervised}     & 47.0          & 63.1          & 73.8          & 71.1          & 73.4          & 4.73          & 5.33          & 4.30          & 6.97          & 4.56          \\ \hline
UDA-VAE++    & \textbf{51.4} & \textbf{65.9} & \textbf{76.5}          & \textbf{73.0}          & \textbf{75.5} & \textbf{3.88} & \textbf{5.23} & \textbf{3.78}          & \textbf{6.25}          & \textbf{4.06} \\
\hline
\end{tabular}
\caption{Unsupervised Domain Adaptation for MM-WHS Dataset from \textbf{CT to MRI}. The best score for Dice$\uparrow$ and ASSD$\downarrow$ are in \textbf{bold}.}
\label{CT_MRI_Dice_ASST}
\end{table*}

Segmentation image visual details
\begin{figure*}
    \centering
    \includegraphics[width=2.65cm]{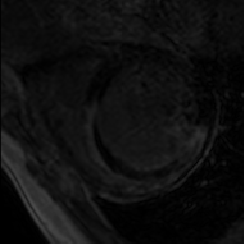}
    \includegraphics[width=2.65cm]{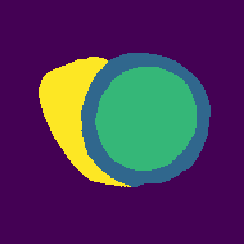}
    \includegraphics[width=2.65cm]{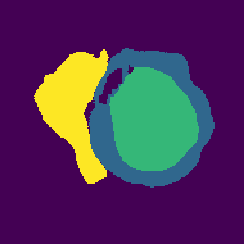}
    \includegraphics[width=2.65cm]{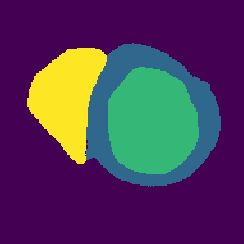}
    \includegraphics[width=2.65cm]{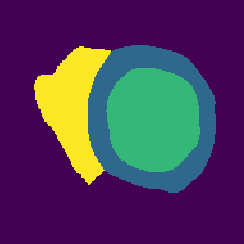}
    \\
    \includegraphics[width=2.65cm]{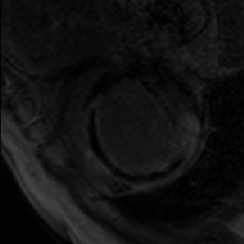}
    \includegraphics[width=2.65cm]{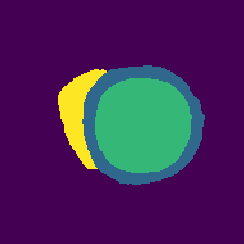}
    \includegraphics[width=2.65cm]{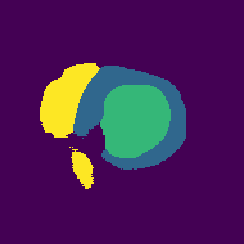}
    \includegraphics[width=2.65cm]{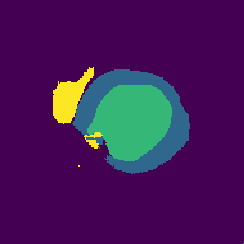}
    \includegraphics[width=2.65cm]{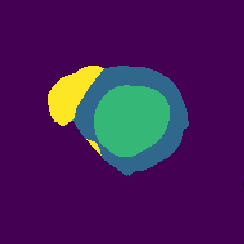}
    \caption{Segmentation output from MS-CMRSeg Dataset (bSSFP to LGE-MRI). From left to right: LGE-MRI, Ground truth, CFDNet\cite{wu2020cf}, UDA-VAE\cite{wu2021unsupervised}, UDA-VAE++. For the segmentation, we use yellow, green, and dark green to represent RV, MYO, and LV, respectively.}
    \label{Visual_Seg}
\end{figure*}

\begin{figure*}
    \centering
    \includegraphics[width=2cm]{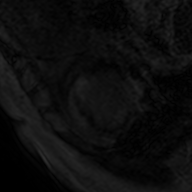}
    \includegraphics[width=2cm]{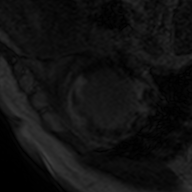}
    \includegraphics[width=2cm]{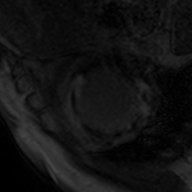}
    \includegraphics[width=2cm]{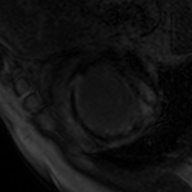}
    \includegraphics[width=2cm]{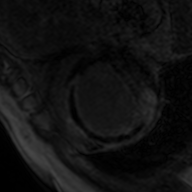}
    \includegraphics[width=2cm]{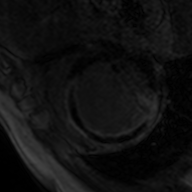}
    \includegraphics[width=2cm]{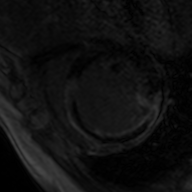}
    \includegraphics[width=2cm]{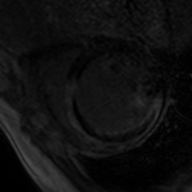} \\
    \includegraphics[width=2cm]{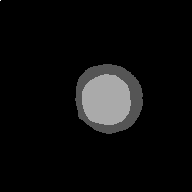}
    \includegraphics[width=2cm]{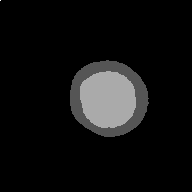}
    \includegraphics[width=2cm]{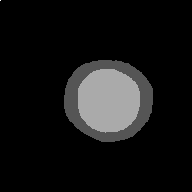}
    \includegraphics[width=2cm]{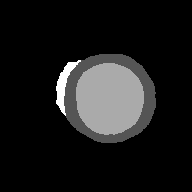}
    \includegraphics[width=2cm]{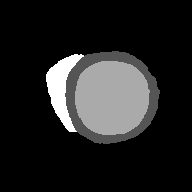}
    \includegraphics[width=2cm]{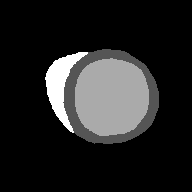}
    \includegraphics[width=2cm]{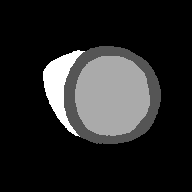}
    \includegraphics[width=2cm]{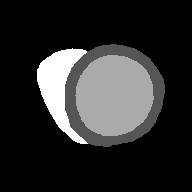} \\
    \includegraphics[width=2cm]{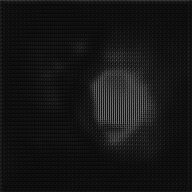}
    \includegraphics[width=2cm]{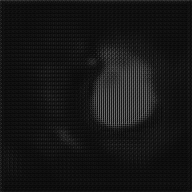}
    \includegraphics[width=2cm]{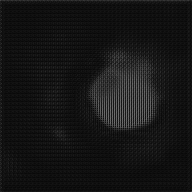}
    \includegraphics[width=2cm]{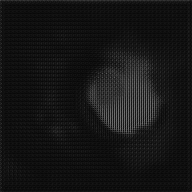}
    \includegraphics[width=2cm]{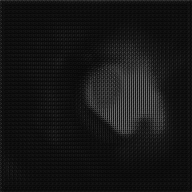}
    \includegraphics[width=2cm]{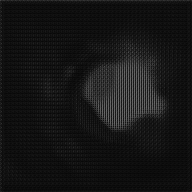}
    \includegraphics[width=2cm]{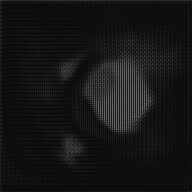}
    \includegraphics[width=2cm]{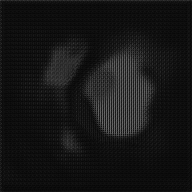} \\
    \includegraphics[width=2cm]{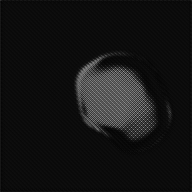}
    \includegraphics[width=2cm]{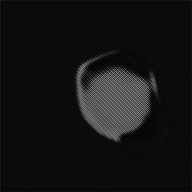}
    \includegraphics[width=2cm]{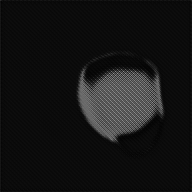}
    \includegraphics[width=2cm]{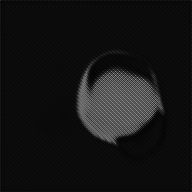}
    \includegraphics[width=2cm]{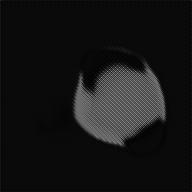}
    \includegraphics[width=2cm]{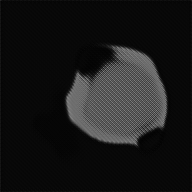}
    \includegraphics[width=2cm]{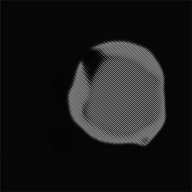}
    \includegraphics[width=2cm]{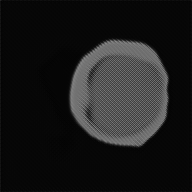} \\
    \caption{Reconstruction Images from MS-CMRSeg Dataset (LGE-MRI). From top to bottom row: LGE-MRI images, corresponding segmentation ground truth, UDA-VAE, UDA-VAE++. }
    \label{Visual_Recon}
\end{figure*}

\subsection{Implementation Details}
We use Adam optimizer \cite{kingma2014adam} and Pytorch framework \cite{paszke2019pytorch} to train our model for 30 epochs. The learning rate is initialized at 1e-4 and is reduced by 10 \% after every epoch. The batch size is 12, which takes about 1 hour to converge on a single NVIDIA Tesla V100 GPU. The network weight follows Xavier initialization \cite{glorot2010understanding}. Neither gradient scaling nor gradient clipping is applied during training.

\subsection{Datasets}

We consider two benchmark datasets for model performance comparison, including Multi-Modality Whole Heart Segmentation (MM-WHS) Challenge dataset \cite{zhuang2016multi} and Multi-Sequence Cardiac MR Segmentation (MS-CMRSeg) Challenge dataset \cite{zhuang2018multivariate}  

\textbf{MM-WHS Dataset} contains 20 labeled CT 3D images and 20 labeled MRI 3D images, which are unpaired. The original size of all images is 240$\times$220, which are cropped with a Region of Interest (ROI) of 192 × 192. 

\textbf{MS-CMRSeg Dataset} contains 35 labeled bSSFP CMR 3D images and 45 labeled 3D LGE-MRI images, which are also not paired. Each image is cropped to a size of 192$\times$192. 

Similar to \cite{wu2020cf, wu2021unsupervised}, we include the following three structures in MS-CMRSeg dataset for segmentation: the myocardial (MYO), the left ventriculus (LV), and the right ventriculus (RV). In MM-WHS Dataset, we include five structures: the myocardial (MYO), the left ventriculus (LV), the right ventriculus (RV), the left atrium blood cavity (LA), and the right atrium blood cavity (RA).  For both datasets, We remove the LGE-MRI ground truth during bSSFP to LGE-MRI experiments, remove the bSSFP ground truth during LGE-MRI to bSSFP, and remove the MRI ground truth during CT to MRI experiments. The train-test split strategy is consistent with \cite{dou2019pnp, chen2020unsupervised, wu2020cf, wu2021unsupervised}

\subsection{Evaluation Metrics}
We use two commonly used evaluation metrics for segmentation, including Dice coefficient (\%) and Average Symmetric Surface Distance (ASSD) (mm). The Dice coefficient calculates the agreement between the predicted segmentation and ground truth segmentation by dividing the intersection area by the total pixels in both images. ASSD measures the segmentation accuracy at boundary-level using the Euclidean
distance of the closest surface voxels between two segmentations \cite{heimann2009comparison}. All metrics are in the format of the mean. A higher Dice and a lower ASSD score indicate better segmentation performances. 

\subsection{Ablation Study}
In this subsection, we investigate the contribution of our model components via an ablation study, using the Dice coefficient as the evaluation metric. Specifically, we gradually add individual components and see how the presence of that component will affect the model performances. 

Table \ref{ablation} shows the quantitative results of the ablation study. It is shown that most proposed modules will improve the Dice scores. For example, sequential reparameterization, adding Attention, Global, and Local MI estimation increases the Dice score. Besides, prior info matching will slightly decrease RV but increase MYO, indicating overall performance improvement. 

\subsection{Qualitative Comparison}

Fig. \ref{Visual_Seg} shows the visual comparison for segmentation among different models, including CFDNet, UDA-VAE, and the proposed UDA-VAE++. It is shown that the proposed UDA-VAE++ leads to the best structure representation, the best edge preservation, and is the closest to the ground truth. In contrast, CFDNet and UDA-VAE have a significant segmentation error between MYO, RV, and the background. 

Fig. \ref{Visual_Recon} displays the visual comparison for reconstruction between different models. Here we only compare UDA-VAE++ with UDA-VAE since UDA-VAE is the only related work that considers image reconstruction. It is shown that the proposed UDA-VAE++ displays significantly better reconstruction than UDA-VAE. UDA-VAE++ has excellent edge preservation, shape representation, and class segmentation. In comparison, UDA-VAE has a significant amount of blurs and artifacts.

\subsection{Quantitative Comparison}

The quantitative comparison utilize several state-of-the-art models, including PnP-AdaNet \cite{dou2019pnp}, SIFA \cite{chen2020unsupervised}, UDA-VAE \cite{wu2021unsupervised}, and the proposed UDA-VAE++.

Table \ref{CT_MRI_Dice_IOU} shows the quantitative comparison for UDA with MS-CMRSeg Dataset (bSSFP to LGE-MRI). We can find that the proposed UDA-VAE++  has the best Dice and ASSD score in terms of MYO and LV segmentation. While SIFA has a slight advantage for RV segmentation, it underperforms our model for all other metrics in the table. Therefore, we can conclude that the proposed UDA-VAE++ has the best performance in this experiment.

Table \ref{MRI_CT_Dice_IOU} shows the quantitative comparison for UDA with MS-CMRSeg Dataset (LGE-MRI to bSSFP). We can observe that the proposed UDA-VAE++ has the best Dice and ASSD score in terms of MYO and LV segmentation. Despite SIFA's success in Dice score at RV segmentation, it significantly underperforms our method for all other metrics. Overall, the proposed UDA-VAE++ has the best result in this comparison.

Table \ref{CT_MRI_Dice_ASST} shows the quantitative comparison for UDA with MM-WHS Dataset (CT to MRI). We can see that the proposed UDA-VAE++ has the best Dice and ASSD score in terms of all segmentations (MYO, LA, LV, RA, RV).

\section{Conclusion}
This paper introduces UDA-VAE++, an unsupervised domain adaptation framework for cardiac segmentation. Through mutual information estimation and maximization, we make the reconstruction and segmentation task mutually beneficial. Moreover, we introduce the sequential reparameterization design, allowing information flow between multi-scale latent space features. Extensive experiments demonstrate that our model achieved state-of-the-art performances on benchmark datasets. Our future work will integrate the proposed mutual information estimation block with self-supervised domain adaptation methods. We also aim to extend our framework to other medical image segmentation tasks (e.g., brain image segmentation).
\section{Acknowledgement}
We appreciate Dr. Fuping Wu and Zirui Wang's kind help with mathematical deduction understanding and model workflow advice.

{\small
\bibliographystyle{ieee_fullname}
\bibliography{egbib}
}
\clearpage

\section{Supplementary Material}

We use the notation of Sec.~3.3.

\begin{theorem}[Structure-preserving decomposition]
\label{thm:main}
\hfill\\
(i) $\mathbb{E}_{p_{D}}[\log p_{\theta}(x,y)]=\mathbb{E}_{p_{D}}[\mathrm{LB}]+\Delta$, where
\[
\Delta=\mathbb{E}_{p_{D}}D_{KL}\left(q_{\phi}(z\mid x)\,\|\,p_{\theta}(z\mid x,y)\right)\geq 0.
\]
(ii) $\mathbb{E}_{p_{D}}[\mathrm{LB}]=-\mathcal{E}-I(x;z)-D_{KL}\left(q(z)\,\|\,p(z)\right)$.

\noindent(iii) For every $\lambda\geq 0$,
\[
\begin{aligned}
\mathbb{E}_{p_{D}}[\mathrm{LB}]+\lambda I(x;\hat{y})
&=-\mathcal{E}-D_{KL}\left(q(z)\,\|\,p(z)\right)\\
&\quad-I(x;z\mid\hat{y})-(1-\lambda)\,I(x;\hat{y}).
\end{aligned}
\]
\end{theorem}

\begin{lemma}
\label{lem:bridge}
For any reconstruction $R$ with $I(\hat{y};R)$, $I(\hat{y};R\mid x)$ and $I(\hat{y};x\mid R)$ finite,
\[
I(x;\hat{y})=I(\hat{y};R)-I(\hat{y};R\mid x)+I(\hat{y};x\mid R).
\]
\end{lemma}

\noindent\textbf{Proof of Theorem~\ref{thm:main}.}
Recall that $q(x,y,z)=p_{D}(x,y)\,q_{\phi}(z\mid x)$, $q(z)=\mathbb{E}_{p_{D}}[q_{\phi}(z\mid x)]$, and $p_{\theta}(x,y,z)=p(z)\,p_{\theta}(y\mid z)\,p_{\theta}(x\mid y,z)$.

\noindent\textbf{(i)} By the chain rule of $p_{\theta}$, the three terms of Eq.~\ref{lb} combine into
\begin{equation}
\mathrm{LB}(x,y)=\mathbb{E}_{q_{\phi}(z\mid x)}\left[\log p_{\theta}(x,y,z)-\log q_{\phi}(z\mid x)\right].
\end{equation}
With $p_{\theta}(z\mid x,y)=p_{\theta}(x,y,z)/p_{\theta}(x,y)$,
\begin{equation}
\begin{aligned}
\log p_{\theta}(x,y)-\mathrm{LB}(x,y)
&=\mathbb{E}_{q_{\phi}(z\mid x)}\left[\log\frac{q_{\phi}(z\mid x)}{p_{\theta}(z\mid x,y)}\right]\\
&=D_{KL}\left(q_{\phi}(z\mid x)\,\|\,p_{\theta}(z\mid x,y)\right)\geq 0 .
\end{aligned}
\end{equation}
Taking the expectation over $p_{D}$ gives (i).

\noindent\textbf{(ii)} The last two terms of Eq.~\ref{lb} give $-\mathcal{E}$ after the expectation over $p_{D}$, so
\begin{equation}
\mathbb{E}_{p_{D}}[\mathrm{LB}]=-\mathcal{E}-\mathbb{E}_{p_{D}}D_{KL}\left(q_{\phi}(z\mid x)\,\|\,p(z)\right).
\end{equation}
Inserting the aggregate posterior $q(z)$,
\begin{equation}
\begin{aligned}
&\mathbb{E}_{p_{D}}D_{KL}\left(q_{\phi}(z\mid x)\,\|\,p(z)\right)\\
=&\,\mathbb{E}_{q}\left[\log\frac{q_{\phi}(z\mid x)}{q(z)}\right]+\mathbb{E}_{q(z)}\left[\log\frac{q(z)}{p(z)}\right]\\
=&\,I(x;z)+D_{KL}\left(q(z)\,\|\,p(z)\right).
\end{aligned}
\end{equation}
The first term is $I(x;z)$ because the $(x,z)$-marginal of $q$ is $p_{D}(x)\,q_{\phi}(z\mid x)$. Since the encoder only reads $x$, $I((x,y);z)=I(x;z)$.

\noindent\textbf{(iii)} The chain rule of mutual information gives
\begin{equation}
I(x;z,\hat{y})=I(x;z)+I(x;\hat{y}\mid z)=I(x;\hat{y})+I(x;z\mid\hat{y}).
\end{equation}
As $\hat{y}$ is computed from $z$, $x\to z\to\hat{y}$ is a Markov chain and $I(x;\hat{y}\mid z)=0$. Hence
\begin{equation}
I(x;z)=I(x;\hat{y})+I(x;z\mid\hat{y}).
\end{equation}
Substituting this into (ii) and adding $\lambda I(x;\hat{y})$ gives (iii).

\noindent\textbf{Proof of Lemma~\ref{lem:bridge}.}
The chain rule of mutual information gives
\begin{equation}
I(\hat{y};x,R)=I(\hat{y};R)+I(\hat{y};x\mid R)=I(\hat{y};x)+I(\hat{y};R\mid x).
\end{equation}
Since $I(\hat{y};R)$ and $I(\hat{y};x\mid R)$ are finite, $I(\hat{y};x,R)$ is finite, and so are $I(\hat{y};x)$ and $I(\hat{y};R\mid x)$. Rearranging the two expansions gives the identity.

\end{document}